\documentclass[final]{aipproc}
\layoutstyle{6x9}
\usepackage{amsmath}
\begin{document}
\title{Kinetic Theory of Random Graphs}
\author{E.~Ben-Naim}{address={Theoretical Division and Center for
Nonlinear Studies, Los Alamos National Laboratory, Los Alamos, NM
87545}}
\author{P.~L.~Krapivsky}{address={
Center for Polymer Studies and
Department of Physics, Boston University, Boston, MA
02215}}

\begin{abstract}

Statistical properties of evolving random graphs are analyzed using
kinetic theory. Treating the linking process dynamically, structural
characteristics of links, paths, cycles, and components are obtained
analytically using the rate equation approach. Scaling laws for finite
systems are derived using extreme statistics and scaling arguments. 

\end{abstract}

\maketitle

\section{Introduction}

Random graphs have been studied in many disciplines including 
statistical physics, chemical physics, combinatorics, probability
theory, and computer science \cite{sr,er,bb,jlr,jklp,pjf1}. For
example, they are used to model percolation in polymerization
processes \cite{pjf,whs} and phase transitions in algorithmic
complexity \cite{bbckw}.

A random graph is a set of nodes that are joined by random links. When
the number of links exceeds a threshold, a connected component
containing a finite fraction of all nodes, the giant component,
emerges. Essentially, random graphs are a mean-field model of
percolation \cite{ds,kcbh}.

In this short review, we summarize our recent work on random graphs
\cite{bk1,bk2}. We describe how, by treating the linking process
dynamically, random graphs can be studied using kinetic
theory. Structures such as paths, cycles, and components grow via
elementary aggregation processes and their distributions can be
obtained using the rate equation approach
\cite{mvs,chandra,da,fl,jbm,hez}. This technique complements the
combinatorial methods, traditionally used to analyze random graphs
\cite{bb,jlr,jklp}.

\section{The evolving random graph}

A graph is a collection of nodes that are joined by links, and in a
random graph, the links are random.  There are different types of
graphs. In a static graph, the links are generated instantaneously,
while in an evolving graph the links are generated sequentially. In a
simple graph, a given pair of nodes may be connected by a single link
only, but in a multigraph they may be connected by multiple links.

We consider the following random graph model.  Starting with no links
and $N$ disconnected nodes, links are sequentially added between
randomly selected pairs of nodes. This linking process continues ad
infinitum with constant rate, set to $(2N)^{-1}$ without loss of
generality. The two nodes selected for linking may not be different,
and additionally, the number of links between two nodes is not
limited. In other words, we consider a random evolving
multi-graph. Additionally, we consider the infinite system size limit,
$N\to \infty$, where statistical fluctuations can be usually ignored.

\section{Links, Paths, and Cycles} 

At time $t$, the total number of links is on average $Nt/2$, and
therefore, the average number of links per node (the degree) equals
time $t$. Let $l$ be the degree of a node. It undergoes the additive
stochastic process $l\to l+1$.  The probability $F_l$ that the degree
of a node equals $l$, the degree distribution, satisfies
\begin{equation}
\label{fl-eq}
\frac{dF_l}{dt}=F_{l-1}-F_l
\end{equation}
with the initial condition $F_l(0)=\delta_{l,0}$. Therefore, the
degree distribution is Poissonian
\begin{equation}
\label{fl}
F_l=\frac{t^l}{l!}\,e^{-lt}
\end{equation}
with the mean degree equal to time, $\langle l\rangle =t$. 

A pair of nodes may be connected by a consecutive series of links
 forming a path. When a newly added link connects two paths of lengths
 $n$ and $m$, a longer path is formed: $n,m \to n+m+1$.  Thus, paths
 undergo an aggregation process. Let $P_l(t)$ be the density of {\it
 distinct} paths containing $l$ links at time $t$. This density
 satisfies
\begin{equation}
\label{pl-eq}
\frac{dP_l}{dt}=\sum_{n+m=l-1}P_nP_m
\end{equation}
for $l>0$ and $P_0(t)=1$. The initial condition is
$P_l(0)=\delta_{l,0}$. Therefore, the path length density is
\begin{equation}
\label{pl}
P_l=t^l.
\end{equation}
For example, the first quantity $P_1=t$ reflects that the link density
is equal to $t/2$ and that every link corresponds to two distinct
paths of length one. The total density of paths $P_{\rm
tot}\equiv\sum_l P_l$ diverges at the percolation time, $P_{\rm
tot}=(1-t)^{-1}$ as $t\to1$. At this time, the system develops a giant
component that eventually percolates through the entire system. The
divergence of the total number of paths is typical: average quantities
as well as typical characteristics diverge near the transition point.
For example, the typical path length diverges as the percolation time is
approached
\begin{equation}
\label{l-typical}
l\sim (1-t)^{-1},
\end{equation}
as seen from the average path length, $\langle l\rangle =\sum_l l
P_l/\sum_l P_l=t(1-t)^{-1}$.

When two nodes along a path are linked, a cycle forms. Cycles have
been studied extensively \cite{mm,sj,sj1,bjk} and for example,
they are useful for characterizing phase transitions in algorithmic
complexity \cite{bbckw}. Let the average number of cycles of size $l$
at time $t$ be $Q_l(t)$. It is coupled to the path length density via
the rate equation
\begin{equation}
\label{ql-eq}
\frac{dQ_l}{dt}=\frac{1}{2}\,P_{l-1}.
\end{equation}
The right-hand side equals the link creation rate $1/(2N)$ times the
total number of paths $NP_{l-1}$.  As a result, the cycle length
distribution is
\begin{equation}
\label{ql}
Q_l=\frac{t^l}{2l}.
\end{equation}
Thus, at the percolation time, the cycle length distribution is
inversely proportional to the length, $Q_l(t=1)=(2l)^{-1}$. In
general, size distributions decay exponentially away from the
percolation point and algebraically precisely at the percolation
point. The total number of cycles in the system $Q_{\rm
tot}\equiv\sum_l Q_l$ is $Q_{\rm tot}=\frac{1}{2}\ln
\frac{1}{1-t}$. It weakly diverges as the percolation point is
approached. We note that the average number of cycles is not an
extensive quantity and for large systems, it saturates at a finite
value. The number of cycles is therefore a non-self-averaging
quantity, it fluctuates from realization to realization.

What is the probability $S(t)$ that the system contains no cycles at
time $t$? Since the cycle formation process is random, and the cycle
production rate is $J=dQ_{\rm tot}/dt=\frac{1}{2(1-t)}$, then
$dS/dt=-S\,J$ or alternatively, 
\begin{equation}
\label{st-eq}
\frac{dS}{dt}=-\frac{S}{2(1-t)}.
\end{equation}
Therefore, the probability that the system contains no cycles decays
with time as follows
\begin{equation}
\label{st}
S=(1-t)^{1/2}.
\end{equation}
This survival probability shows that the system is bound to nucleate
at least a single cycle prior to the percolation time. 

Following similar reasoning, properties of the first cycle may be
obtained.  The size distribution of the first cycle $G_l$ obeys a
simple generalization of (\ref{ql-eq})
\begin{equation}
\label{gl-eq}
\frac{dG_l}{dt}=\frac{1}{2}S P_{l-1}. 
\end{equation}
The rate by which the first cycle is produced is simply the rate by
which all cycles are produced times the probability that there are no
cycles in the system. Since the first cycle must be produced by the
percolation time, the final size distribution of the first cycle is
$G_l(1)=\frac{1}{2}\int_0^1 dt\, SP_{l-1}$ and performing the
integration gives
\begin{equation}
\label{gl}
G_l(t=1)=\frac{\Gamma(3/2)\Gamma(l)}{2\Gamma(l+3/2)}.
\end{equation}
This size distribution has an algebraic tail, $G_l(1)\sim l^{-3/2}$
for $l\gg 1$. The characteristic exponent for the tail of $G_l$ is
larger than the characteristic exponent for the tail of $Q_l$: the
first cycle is created earlier, and thus, it must be smaller.

\section{Finite Components: size distribution}

A component is a set of connected nodes: every pair of nodes in a
component is connected by a path.  Components merge due to linking: a
link placed between two distinct components causes the two to
join. There are $i\times j$ ways to join disconnected components of
size $i$ and $j$ and hence, components undergo the aggregation process
$(i,j) \to i+j$ with the aggregation rate $ij/(2N)$.

Let $c_k(t)$ be the density of components containing $k$ nodes at time
$t$. The component size distribution obeys the nonlinear rate equation
\begin{equation}
\label{ck-eq}
\frac{dc_k}{dt}=\frac{1}{2}\sum_{i+j=k}(ic_i)(jc_j)-k\,c_k.
\end{equation}
The initial condition is $c_k(0)=\delta_{k,1}$.  The gain term
represents merger between two components whose sizes sum up to $k$ and
the loss term accounts for links involving a node inside a component
of size $k$.

The moments of the size distribution, $M_n=\sum_k k^n c_k$, provide a
 useful probe of the dynamics. For example, the second moment obeys
 the closed equation $dM_2/dt=M_2^2$ and with the initial condition
 $M_2(0)=1$, the solution is
\begin{equation}
\label{m2}
M_2=(1-t)^{-1},
\end{equation}
for $t<1$. The divergence shows that the system undergoes a
percolation transition. In a finite time $t_g=1$, an infinite
component, the giant component, is formed. Past the percolation point,
the giant component contains a finite fraction of the nodes, and
eventually, it grows to engulf the entire system.

The component size distribution can be obtained
analytically\footnote{A convenient solution method is as follows. The
time dependence is ``peeled'' first, $c_k=C_kt^{k-1}e^{-kt}$ with the
coefficients satisfying $(k-1)C_k=\sum_{i+j=k} (iC_i)(jC_j)$. The
generating function $G(z)=\sum_k kC_k e^{kz}$ satisfies the
differential equation \hbox{$(1-G)G'=G$} and consequently
$Ge^{-G}=e^z$. The coefficients are found using the Lagrange inversion
formula \cite{bk2}. The very same technique can be used to derive the
joint distributions described in the next section.}
\begin{equation}
\label{ck}
c_k(t)=\frac{k^{k-2}}{k!}\,t^{k-1}\,e^{-kt}.
\end{equation}
This size distribution decays exponentially away from the percolation
point and algebraically at the percolation point, $c_k(1)\simeq
(2\pi)^{-1/2}k^{-5/2}$. Both behaviors follow from the scaling
behavior $c_k(t)\to (1-t)^5\Phi_C\big(k(1-t)^2\big)$ with the typical
component size
\begin{equation}
\label{k-typical}
k\sim (1-t)^{-2}
\end{equation}
and the scaling function
$\Phi_C(x)=(2\pi)^{-1/2}x^{-5/2}\exp(-x/2)$. The large-size algebraic
decay of the size distribution is reflected by the small-argument
behavior of the scaling function. Hence, the size distribution
exhibits dynamical scaling in the vicinity of the percolation
transition. This behavior is generic: size distributions obey
dynamical scaling near the percolation point and for example, both the
path length density (\ref{pl}) and the cycle length density (\ref{ql})
can be written in a self-similar form.

Other statistical properties including for example the moments follow
from the generating function, $c(z,t)=\sum_k k\, c_k(t) e^{kz}$, that
can be written explicitly
\begin{equation}
\label{czt}
c(z,t)=t^{-1}G(z+\ln t-t),
\end{equation}
in terms of the auxiliary generating function
\begin{equation}
\label{gz}
G(z)=\sum_{k=1}^{\infty} \frac{k^{k-1}}{k!}e^{kz}.
\end{equation}
This function satisfies $Ge^{-G}=e^z$ or alternatively,
$(1-G)dG/dz=G$. 

Combining the latter relation and the behavior of the generating
function near $z=0$, the second moment result (\ref{m2}) is
generalized to all times, \hbox{$M_2=u/[t(1-u)]$} with $u=G(\ln
t-t)$. This quantity satisfies
\begin{equation}
\label{u}
u\,e^{-u}=t\,e^{-t}.
\end{equation}
Let the fraction of nodes outside finite components be $g=1-M_1$ with
\hbox{$M_1=c(z=0)=u/t$}.  For $t<1$, there is a single solution $u=t$
and therefore all nodes are in finite components, $g=0$. But when
$t>1$, there is an additional nontrivial solution, and as a result,
this fraction becomes finite, $g>0$. In particular, at the late stages
of the process, the giant component contains almost all nodes,
$g(t)\to 1$, and furthermore, since $u\simeq te^{-t}$ then $1-g\simeq
c_1= e^{-t}$, indicating that other than the giant component, there
are only a few isolated nodes. Also, just past the percolation point,
the fraction of nodes outside finite components grows linearly,
$g(t)\simeq 2(t-1)$.

\section{Finite Components: Joint Distributions}

We have seen that components undergo an aggregation process of one
kind and that links, paths, and cycles undergo growth or aggregation
processes of another kind. By combining these separate processes into
a bi-aggregation process \cite{kb} involving two variables, say the
component size and the node degree, a more detailed analysis of
structural properties of finite components is possible.

\subsection{Links}

Each node can be characterized by two indices: its degree $l$ and the
size $k$ of the component it belongs to. The distribution $f_{l,k}$ of
nodes of degree $l$ in components of size $k$ satisfies
\begin{eqnarray}
\label{flk-eq}
\frac{df_{l,k}}{dt}=\sum_{i+j=k}(jc_j)[(i-1)f_{l,i}+f_{l-1,i}]-k\,f_{l,k}.
\end{eqnarray}
The initial conditions are $f_{l,k}(0)=\delta_{l,0}\,\delta_{k,1}$.  The
first gain term accounts for linking events that leave the node degree
unchanged (the added link involves other nodes in the component),
while the second gain term represents linking events that augment the
node degree by one. Of course, the component size distribution and the
node-degree distribution can be obtained by summation of the joint
distribution, so that Eqs.~(\ref{fl-eq}) and (\ref{ck-eq}) can be 
recovered from (\ref{flk-eq}). 

The joint distribution can be obtained analytically 
\begin{equation}
f_{l,k}(t)=\frac{(k-1)^{k-l-2}}{(l-1)!(k-l-1)!}\,t^{k-1}\,e^{-kt}
\end{equation}
for $1\leq l<k$ and $f_{0,k}(t)=\delta_{k,1}\,e^{-t}$ for
$l=0$. Fixing the node degree, the joint distribution decays
algebraically at large sizes at the critical point
$f_{l,k}(t=1)\sim\,k^{-3/2}$. Exponential decay occurs elsewhere. 

The generating function $f(z,w)=\sum_{l,k}e^{kz}w^lf_{l,k}$ is
expressed directly in terms of auxiliary generating function 
\begin{equation}
f(z,w)=e^{wG(z+\ln t-t)+z-t}.
\end{equation}
Average quantities and correlations follow from the 
generating function, and for example, the average node degree, the average 
component size, and the average correlation between the two are 
\begin{eqnarray}
\langle l\rangle=u,\qquad
\langle k\rangle=\frac{1}{1-u},\qquad 
\langle kl\rangle=\frac{2u}{1-u}.
\end{eqnarray}
Below the percolation transition, the average node degree equals time,
$\langle l\rangle =t$. Above the percolation transition, the average node
degree is reduced $\langle l\rangle <t$ because nodes in finite
components have less connections than the rest of the nodes. The
average degree vanishes in the long time limit $\langle l\rangle
\simeq te^{-t}$. In comparison, the fraction of isolated nodes is
$c_1\simeq e^{-t}$.  Interestingly, the properly normalized
correlation between the node degree and the component size is time
independent, $\langle kl\rangle/\langle k\rangle\langle l\rangle=2$.
The node degree and the component size are correlated: nodes that have
more links likely belong to larger components.

\subsection{Paths}

Since every two nodes in a component are connected, there must be a
path connecting them. Let $p_{l,k}$ be the density of paths of length
$l$ in components of size $k$. There is the obvious bound $0\leq l\leq
k-1$ and additionally, there is a sum rule $\sum_l p_{l,k}=k^2 c_k$
reflecting that there are $k^2$ distinct paths in a component of size
$k$.  The density of linkless paths is $p_{0,k}=kc_k$.

The path length and the component size separately undergo an
aggregation process and combining the two processes, these two indices
undergo a bi-aggregation process.  The joint distribution evolves
according to the rate equation
\begin{equation}
\label{plk-eq}
\frac{dp_{l,k}}{dt}=\!\!\!\sum_{\substack{i+j=k\\n+m=l-1}}\!\!\!p_{n,i}p_{m,j}
+\sum_{i+j=k}(ip_{l,i})(jc_j)-kp_{l,k}.
\end{equation}
The initial conditions are $p_{l,k}(0)=\delta_{k,1}\delta_{l,0}$.
There are two separate convolutions: one over the path length and one
over the component size. The first term on the right-hand side of
Eq.~(\ref{plk-eq}) describes newly formed paths due to linking and the
last two terms correspond to paths that do not contain the newly
placed link.

The path length density is 
\begin{equation}
\label{plk}
p_{l,k}=(l+1)\,\frac{k^{k-l-2}}{(k-l-1)!}\,t^{k-1}\,e^{-kt}.
\end{equation}
The two shortest paths satisfy $p_{0,k}=kc_k$ and
$p_{1,k}=2(k-1)c_k$. The latter reflects that there are $k-1$ links in
a (tree) component of size $k$. Also, the longest possible path,
$l=k-1$, corresponds to a linear (chain-like) component, and the
density of such components, $p_{k-1,k}=t^{k-1}e^{-kt}$, decays
exponentially with length, so that such components are typically
small.

The path length density can be simplified in the limit $k\gg l\gg 1$, 
\begin{equation}
\label{plk-largek}
p_{l,k}\simeq l\,(2\pi k^3)^{-1/2}\,t^{k-1}\,e^{k(1-t)}\,e^{-l^2/2k}.
\end{equation}
As was the case for the component size distribution, the path
length density is self-similar in the vicinity of the percolation
point, $p_{l,k}\to (1-t)^2 \Phi_p\left(k(1-t)^2,l(1-t)\right)$,
with the scaling function
\begin{equation}
\label{phip}
\Phi_p(x,y)=y\,(2\pi x^3)^{-1/2}\exp(-y^2/2x).
\end{equation}
The characteristic path length is as in (\ref{l-typical}) and the
characteristic component size is as in (\ref{k-typical}).  At the
percolation point, the path length density (\ref{plk-largek}) is
governed by the factor $\exp(-l^2/2k)$ and therefore, the typical path
length scales as square root of the component size
\begin{equation}
\label{lk}
l\sim k^{1/2}.
\end{equation}

The generating function $p(z,w)=\sum_{l,k} e^{kz} w^l p_{l,k}$ is
  expressed in terms of the auxiliary function (\ref{gz})
\begin{equation}
\label{pzw}
p(z,w)=t^{-1}\frac{G(z+\ln t-t)}{1-wG(z+\ln  t-t)}.
\end{equation}
The total density of paths in finite components \hbox{$p_{\rm
tot}\equiv \sum_{l,k} p_{l,k}$} is therefore \hbox{$p_{\rm
tot}=u/t(1-u)$} and for $t<1$ we recover $p_{\rm tot}=1/(1-t)$.
Expanding $p(z,w)$ in powers of $w$, the total number of paths of
length $l$, $p_l \equiv \sum_k p_{l,k}$, is given by
$p_l=t^{-1}u^{l+1}$ with $u$ satisfying (\ref{u}), in accord with
(\ref{pl}) for $t<1$.

\subsection{Cycles}

We have seen that the cycle length distribution is coupled to  the
path length distribution. In a similar way, the joint distribution of
cycles in finite components of a given size is coupled to the joint
distribution of paths of a given length in components of a given size.

To characterize cycles in a given component size, we consider the
joint distribution $q_{l,k}$, the average number of 
components of size $k$ containing a cycle of length $l$ with $1\leq
l\leq k$. This joint distribution evolves according to the 
linear rate equation
\begin{equation}
\label{ulk-re}
\frac{dq_{l,k}}{dt}=\frac{1}{2}p_{l-1,k}+
\sum_{i+j=k}(iq_{l,i})\,(jc_j)-k\,q_{l,k}
\end{equation}
for $l\geq 1$. Initially, there are no cycles, and therefore
$q_{l,k}(0)=0$. The first term on the right hand side represents
generation of cycles from paths, and the next two terms represent
merger events where only the component size changes.

The joint cycle-length component-size distribution is 
\begin{equation}
\label{ulkt}
q_{l,k}(t)=\frac{1}{2}\,\frac{k^{k-l-1}}{(k-l)!}t^ke^{-kt}.
\end{equation}
The smallest cycle, $l=1$, is a self-connection, and the average
number of such cycles is $q_{1,k}=\frac{t}{2}\,kc_k$. The largest
cycles are rings, $l=k$, and their total number is on average
$q_{k,k}=\frac{1}{2k}\,t^k\,e^{-kt}$. As for linear chains, the number
of rings decays exponentially with length.

The large-$k$ behavior of the cycle length distribution is similar 
to (\ref{plk-largek})
\begin{equation}
\label{qlk-largek}
q_{l,k}(t)\simeq (8\pi k^3)^{-1/2}\,t^k\,e^{k(1-t)}\,e^{-l^2/2k}.
\end{equation}
This distribution is self-similar in the vicinity of the percolation
transition, \hbox{$q_{l,k}(t)\to
(1-t)^3\Phi_q\left(k(1-t)^2,l(1-t)\right)$}, with the scaling function
$\Phi_q(x,y)=(8\pi x^3)^{-1/2}\exp(-y^2/2x)$.  We see that the cycle
length is characterized by the same scale as the path length, $l\sim
(1-t)^{-1}$.  At the percolation point, the cycle length distribution
(\ref{qlk-largek}) is dominated by the factor $\exp(-l^2/2k)$ so that
when the component size is fixed, the typical cycle length behaves as
the typical path length, $l\sim k^{1/2}$. Moreover, the size
distribution of finite components containing a cycle, $q_k=\sum_l
q_{l,k}$, decays as a power-law at the percolation point, $q_k\simeq
(4k)^{-1}$.

The joint generating function, $q(z,w)=\sum_{l,k}e^{kz}\,w^l\,
q_{l,k}$, is 
\begin{equation}
\label{uzw}
q(z,w)=\frac{1}{2}\ln \frac{1}{1-wG(z+\ln t-t)}.
\end{equation}
As for paths, statistics of cycles are directly coupled to statistics
of components via the generating function $G(z)$.  The total number of
cycle-containing components of finite-size, $q_{\rm tot}=\sum_{l,k}
q_{l,k}$, is therefore $q_{\rm tot}(t)=\frac{1}{2}\,\ln
\frac{1}{1-u}$.  Below the percolation point, $q_{\rm
tot}(t)=\frac{1}{2}\ln \frac{1}{1-t}$, for $t<1$. Moreover, expanding
$q(z,w)$ in powers of $w$ shows that the the cycle length distribution
(in finite components only) is $Q_l=u^l/2l$, in agreement with
(\ref{ql}) prior to the percolation time ($t<1$).

\section{Finite Graphs}

Thus far, we used rate equations to describe infinite systems. While
the rate equation approach can be extended to finite systems, the
resulting equations are difficult to handle \cite{aal,ve}.  When the
number of nodes is finite, fluctuations are no longer negligible, and
instead of a deterministic rate equation approach, a stochastic
approach is needed. Finite-size scaling laws can be conveniently
obtained by combining the exact infinite system results with scaling
and extreme statistics arguments.

Finite-size scaling laws of random graphs are quite interesting. For
example, the giant component nucleates at a size that it much smaller
than the system size. The size of the largest component in the system,
$M$, can be estimated by employing the cumulative component size
distribution and the extreme statistics criterion, $N\sum_{k\geq
M}c_k(t=1)\sim 1$.  Using $c_k\sim k^{-5/2}$ gives
\begin{equation}
\label{M}
M\sim N^{2/3}.
\end{equation}
The largest component in the system grows sub-linearly with the system
size \cite{bb}. This component nucleates very close to the percolation
time. The time $\tau$ when this component emerges approaches unity for
large enough systems as implied by the diverging characteristic size
scale $M\sim (1-\tau)^{-2}$, so that 
\begin{equation}
\label{tau}
1-\tau\sim N^{-1/3}.
\end{equation}
Just past the percolation time, the size of the giant component grows
linearly with time. 

For finite systems, the scaling laws for the typical path length
(\ref{l-typical}) combined with the characteristic component size
(\ref{M}) yields a scaling law for the characteristic path (and cycle)
length
\begin{equation}
\label{ln}
l\sim N^{1/3}.
\end{equation}
One can deduce several other scaling laws and finite-size scaling
functions underlying the path length density. For example,
substituting the percolation time (\ref{tau}) into the total number of
paths $P_{\rm tot}=(1-t)^{-1}$ yields the total path density $P_{\rm
tot}\sim N^{1/3}$. Similarly, the total number of cycles at the
percolation point grows logarithmically with the system size, 
$Q_{\rm tot}(N)\simeq \frac{1}{6}\ln N$.

In finite systems, it is possible that no cycle are created by the
percolation time. This probability decreases algebraically with the
system size, as seen from (\ref{st}) and (\ref{tau})
\begin{equation}
\label{sn}
S\sim N^{-1/6}.
\end{equation}
Moreover, combining the size distribution of the first cycle,
$G_l(1)\sim l^{-3/2}$ with the characteristic cycle scale $l\sim
N^{1/3}$ yields the moments of the size distribution corresponding to
the first cycle
\begin{equation}
\label{ln1}
\langle l^n\rangle \sim N^{n/3-1/6}.
\end{equation}
In particular, the average size of the first cycle is much smaller
than the characteristic cycle length $\langle l\rangle \sim
N^{1/6}$. 

\section{Summary}

In summary, we used kinetic theory to describe structural properties
of random graphs including paths, cycles, and components.  Modeling
the linking process dynamically shows that paths and components
undergo separate aggregation processes. Cycles are generated by paths
and thus, the cycle length distribution is coupled to the path length
distribution. 

Generally, size distributions decay exponentially away from the
percolation point, but at the percolation point, algebraic tails
emerge. As the system approaches this critical point, the size
distributions follow a self-similar behavior and they are
characterized by diverging size scales.

The kinetic theory approach is well-suited for treating infinite
systems. Nevertheless, the behavior of finite systems can be obtained
from heuristic scaling and extreme statistics arguments. This yields
scaling laws for the typical component size, path length, and cycle
length at the percolation point.

The rate equation approach is powerful in that it utilizes a
continuous time variable and therefore, differential, rather than
difference equations. It has been successfully used to model growing
random networks and it should be applicable to more complex random
structures.

\begin{theacknowledgments}
This research was supported by the DOE (W-7405-ENG-36).
\end{theacknowledgments}

\bibliographystyle{aipproc}

\appendix

\end{document}